\documentclass[
    twocolumn,
	prd,
	amssymb,
preprintnumbers,superscriptaddress,
	nofootinbib]{revtex4-2}

\pdfoutput=1
\usepackage{paralist}
\usepackage{graphicx}
\usepackage{enumitem}
\usepackage{latexsym}
\usepackage{amsfonts}
\usepackage{amssymb}
\usepackage{xcolor}
\usepackage[export]{adjustbox}
\usepackage{amsmath}
\usepackage[thinlines]{easytable}
\usepackage{slashed}
\usepackage{dcolumn}
\usepackage{verbatim}
\usepackage{float}
\usepackage{multirow}
\usepackage{booktabs}
\usepackage{xspace}
\usepackage[normalem]{ulem}
\usepackage[
pdfauthor={Jeremy Sakstein}]{hyperref}
\usepackage{tabularx}
\usepackage{lettrine}
\usepackage{setspace}

\newcommand{\LCDM}{$\Lambda$CDM}
\newcommand{\nodata}{$\cdots$}

\newcommand{\mpl}{M_{\rm Pl}}
\newcommand{\dd}{\mathrm{d}}

\begin{document}

\title{DESI Constraints on Exponential Quintessence}
\author{Omar F.~Ramadan}
\email{oramadan@hawaii.edu }
\affiliation{%
Department of Physics $\&$ Astronomy, University of Hawai‘i,
Watanabe Hall, 2505 Correa Road, Honolulu, HI, 96822, USA}
\author{Jeremy Sakstein}
\email{sakstein@hawaii.edu}
\affiliation{%
Department of Physics $\&$ Astronomy, University of Hawai‘i,
Watanabe Hall, 2505 Correa Road, Honolulu, HI, 96822, USA}
\author{David Rubin}
\email{drubin@hawaii.edu}
\affiliation{%
Department of Physics $\&$ Astronomy, University of Hawai‘i,
Watanabe Hall, 2505 Correa Road, Honolulu, HI, 96822, USA}

\date{\today}

\begin{abstract}
The DESI collaboration have recently analyzed their first year of data, finding a preference for thawing dark energy scenarios when using parameterized equations of state for dark energy.~We investigate whether this preference persists when the data is analyzed within the context of a well-studied field theory model of thawing dark energy, exponential quintessence.~No  preference for this model over $\Lambda$CDM is found,  and both models are poorer fits to the data than the Chevallier-Polarski-Linder $w_0$--$w_a$ parameterization.~We demonstrate that the worse fit is due to a lack of sharp features in the potential that results in a slowly-evolving dark energy equation of state that does not have enough freedom to simultaneously fit the combination of the supernovae, DESI, and cosmic microwave background data.~Our analysis provides guidance for constructing dynamical dark energy models that are able to better accommodate the data.
\end{abstract}
\maketitle

The origin of the present day acceleration of the cosmic expansion, \textit{dark energy} (DE), remains a mystery, even after a quarter of a century of research.~Previously, all observations were compatible with dark energy driven by a cosmological constant $\Lambda$, but this has recently been challenged by the DESI first year data release \cite{DESI:2024mwx}, which, when analyzed in combination with the Planck and ACT cosmic microwave background (CMB) measurements and Type Ia supernovae data, either PantheonPlus \cite{Brout:2022vxf}, Union3 \cite{Rubin:2023ovl}, or DESY5 \cite{DES:2024tys},
shows a preference for \textit{thawing dark energy} at the level of $2.5\sigma$, $3.5\sigma$, and $3.9\sigma$ respectively.~In this scenario, the equation of state (EOS) of dark energy $w(z)$ was frozen at a constant value in the past but recently began to evolve away from this, in contrast to $\Lambda$ which has constant $w(z)=-1$.~The thawing DE preference manifests when the data is fit to the Chevallier-Polarski-Linder $w_0$--$w_a$ parameterization \cite{Chevallier:2000qy,Linder:2002et}, which is a phenomenological relation:
\begin{equation}
    \label{eq:CPL}
    w(z)=w_0+w_a\frac{z}{1+z}
\end{equation}
with $w_0$ and $w_a$ free parameters that are fit to the data.~DESI report $w_0=-0.727 \pm 0.067$ and $w_a=-1.05^{+0.31}_{-0.27}$ using CMB+DESI+DESY5 datasets.~

While parameterizations such as $\eqref{eq:CPL}$ are helpful for characterizing the data and as consistency tests of the null $\Lambda$CDM hypothesis, they do not provide any interpretation of data within the context of fundamental physics, motivating investigations of the degree to which competing microphysical models of dark energy can accommodate the data.~In this work, we explore the implications of the first DESI data release for a quintessence model of thawing dark energy, exponential quintessence.~

In quintessence models \cite{Ratra:1987rm,Caldwell:1997ii,Copeland:1997et,Copeland:2006wr,Bahamonde:2017ize}, dark energy is driven by a scalar field $\phi$ with mass $m$ that is initially frozen at its initial condition by Hubble friction so that $w=-1$ but begins to roll sometime in the recent past when $H\sim m$.~This rolling causes the EOS to deviate from $-1$ with $w\ge-1$.~The specific action we consider is 
\begin{equation}
    \label{eq:action}
    S=\int\dd^4 x\sqrt{-g}\left[\frac{\mpl^2}{2}R(g)-\frac12\partial_\mu\phi\partial^\mu\phi-V(\phi)\right] 
\end{equation}
where matter is minimally coupled to the metric $g_{\mu\nu}$ and $\mpl^2=(8\pi G)^{-1}$ is the reduced Planck mass.~In a Friedmann-Lema\^{i}tre-Robertson-Walker (FLRW) universe, the scalar behaves as a perfect fluid with density parameter and equation of state
\begin{align}
    \label{eq:phiEOS}
    w_\phi&=\frac{\dot{\phi}^2-2V(\phi)}{\dot{\phi}^2+2V(\phi)}\\\label{eq:phiDensity}
    \Omega_{\phi}&=\frac{\dot{\phi}^2}{6 H^2\mpl^2}+\frac{V(\phi)}{3 H^2\mpl^2}.
\end{align}
The evolution of the scalar is determined by the Klein-Gordon equation
\begin{equation}
    \label{eq:phiKG}
    \ddot{\phi}+3H\dot{\phi}+\frac{\dd V}{\dd\phi}=0.
\end{equation}
Equations \eqref{eq:phiEOS}--\eqref{eq:phiKG} elucidate how quintessence fields can behave as thawing dark energy.~At early times, when $z\gg1$, the field has initial condition $\phi=\phi_i$ with mass $m^2=V''(\phi_i)$.~Provided that $m\ll H$, the friction term $(3H\dot{\phi}\sim H^2\phi)$ will dominate over the restoring term $(V'(\phi)\sim m^2\phi)$ and the field will be frozen at $\phi_i$.~According to \eqref{eq:phiEOS} and \eqref{eq:phiDensity}, the field behaves a cosmological constant with EOS $w_\phi\approx-1$.~As the universe expands, $H$ decreases, reaching $H\sim m$ around $z\sim1$.~At this point, the field begins to roll or \textit{thaw}, gaining kinetic energy so that $w_\phi>-1$.~The current phase of dark energy corresponds to the scalar slowly-rolling down its potential.

The phenomenology of quintessence DE depends upon the choice of potential.~In this work, we will study the \textit{exponential quintessence} model
\begin{equation}
    \label{eq:expPotential}
    V(\phi)= V_0e^{-\lambda\frac{\phi}{\mpl}},
\end{equation}
an archetypal potential that arises generically in beyond the Standard Model theories such as string theory and supergravity \cite{Wetterich:1994bg,Binetruy:1998rz,Barreiro:1999zs}.~Despite the non-linearity of equation~\eqref{eq:phiKG} and the Fridemann equations, the solution space of exponential quintessence is well-understood because the equations can be written in an autonomous form, implying that dynamical systems methods can be used to identify the steady-state solutions \cite{Copeland:1997et,Copeland:2006wr,Bahamonde:2017ize,Ramadan:2023ivw}.~The system of equations admits a dark energy dominated global attractor with $\Omega_\phi=1$ and 
\begin{equation}
    \label{eq:w_DEAttractor}
    w_\phi=-1+\frac{\lambda^2}{3}
\end{equation}
provided that $\lambda<\sqrt{3}$.~The thawing DE scenario can then be realized within this potential as follows.~At early times, the field is frozen such that $w_\phi\approx-1$ but the field thaws and begins to roll to this attractor at $z\sim 1$.~The current phase of thawing DE corresponds to the approach to this attractor.~The attractor cannot be reached at the present day because this would imply a DE-dominated universe in conflict with observations, and would not match the DESI predictions because $w$ is constant at the attractor.~This introduces some sensitivity to the initial conditions.

\begin{figure*}
\centering
\includegraphics[width=\textwidth]{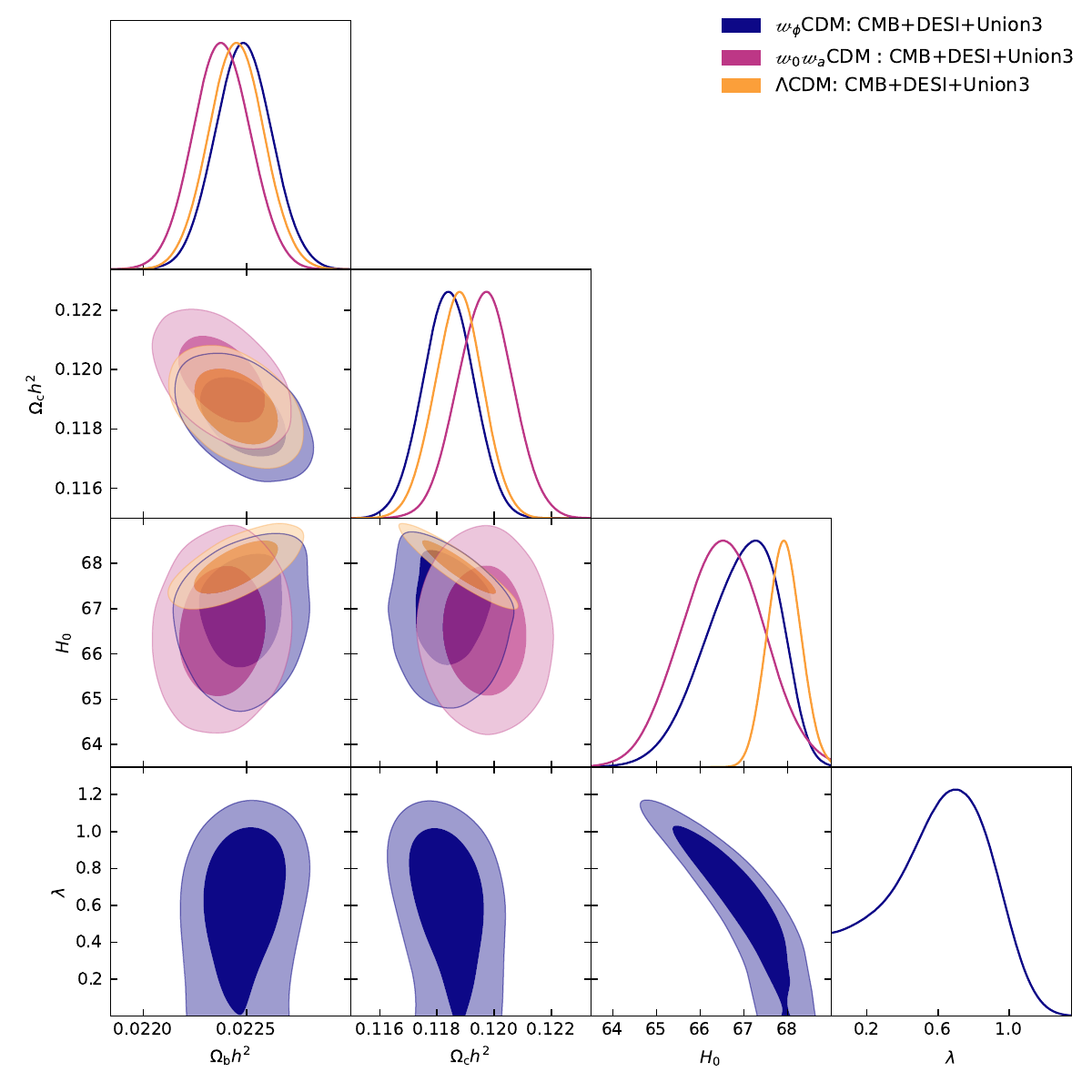}
     \caption{Marginalized posteriors for the cosmological models studied in this work using CMB+DESI+Union3 data.~The inner contours denote the $68\%$ confidence level (CL), while the outer contours denote the $95\%$ CL.~Both $w_0 w_a$CDM and quintessence models encompass the \LCDM\ limit.
     }
\label{fig:contours}
\end{figure*}

We now test this scenario against the DESI data by fitting it to the combination of CMB+DESI+Union3.~The CMB data include Planck 2018 CMB spectra \cite{Planck:2018vyg}, CMB gravitational lensing from a combination of Planck 2020 lensing \cite{Planck:2020olo,Carron:2022eyg} and ACT DR6 \cite{ACT:2023dou,ACT:2023kun}.~This is the same combination of data used by DESI.~We implemented the exponential potential into \texttt{CLASS} \cite{Blas:2011rf,Lesgourgues:2011re} to evolve the cosmology and used  the \texttt{Cobaya} \cite{Torrado:2020dgo} framework to sample using the Markov Chain Monte Carlo (MCMC) \cite{hastings1970monte,metropolis1953equation} algorithm.~Convergence was deemed to be achieved when the standard Gelman-Rubin criteria $R-1<0.01$ \cite{GelmanRubin1992} was met.~To analyze our chains and plotting, we made use of \texttt{GetDist} \cite{Lewis:2019xzd}.~The initial conditions were chosen using the following considerations.~For the initial field, $\phi_i$, we made use of a symmetry of the model:~$\phi\rightarrow\phi+\phi_0$, $V_0\rightarrow V_0\exp({\lambda{\phi_0}/{\mpl}})$ where $\phi_0$ is a constant, which allowed us to fix $\phi_i$ to an arbitrary value without loss of generality.~We chose $\phi_i = -4.583 \mpl$.~For the initial field velocity $\dot{\phi}_i$, we used attractor initial conditions.~At early times, the field is approximately frozen so we assumed slow-roll and set $\ddot{\phi}=0$ in Eq.~\eqref{eq:phiKG} yielding $\dot{\phi_i} = \frac{\lambda V_0}{3H_i}\exp({-\lambda{\phi_i}/{\mpl}})$.~We  modified \texttt{CLASS} to shoot for $V_0$ such that $V_0 = 3 H_0^2 \mpl^2 \Omega_\phi$ in order to close the universe.~We also fit the $w_0$--$w_a$ parameterization to the same data.
\begin{table*}[ht]
\centering
\setlength{\tabcolsep}{9pt}
\def\arraystretch{1.5}
\resizebox{\textwidth}{!}{%
    \small
    \begin{tabular}{lccc}
    \toprule
    \midrule
    Parameter \& Model & {\bf Flat} $\boldsymbol{\Lambda}${\bf CDM} & $\boldsymbol{w_a w_a}${\bf CDM} & $\boldsymbol{w_\phi}${\bf CDM} \\
    \midrule
    \textbf{Sampled Parameters} &&&\\
    $\log(10^{10} A_\mathrm{s})$  & $3.053(3.059)^{+0.013}_{-0.014}$ & $3.040(3.040)\pm 0.013$ & $3.056(3.051)\pm 0.013$ \\
    $n_\mathrm{s}$ & $0.9681(0.9688)\pm 0.0036$ & $0.9657(0.9668)\pm 0.0038$ & $0.9691(0.9692)\pm 0.0037$ \\
    $\Omega_{b} h^2$ & $0.02245(0.02247)\pm 0.00013$ & $0.02238(0.02242)\pm 0.00014$ & $0.02248(0.02249)\pm 0.00014$ \\
    $\Omega_{c} h^2$ & $0.11876(0.11856)\pm 0.00084$ & $0.11968(0.11982)\pm 0.00097$ & $0.11840(0.11839)\pm 0.00089$ \\
    $100\theta_*$ & $1.04199(1.04193)\pm 0.00028$ & $1.04187(1.04185)\pm 0.00029$ & $1.04202(1.04199)\pm 0.00029$ \\
    $\tau_\mathrm{reio}$ & $0.0590(0.0614)\pm 0.0071$ & $0.0526(0.0529)\pm 0.0072$ & $0.0608( 0.0588)^{+0.0070}_{-0.0084}$ \\
    $w_0$ & \nodata & $-0.656(-0.679)\pm 0.099$ & \nodata \\
    $w_a$ & \nodata & $-1.22(-1.14)^{+0.42}_{-0.34}$ & \nodata \\
    $\lambda$ & \nodata & \nodata & $0.60(0.74)^{+0.38}_{-0.27}$ \\
    \midrule
    \textbf{Derived Parameters} &&&\\
    $H_0 \left[\text{km/s/Mpc}\right]$ & $67.92(67.98)\pm 0.39$ & $66.52(66.61)\pm 0.94$ & $66.92(66.613)^{+0.99}_{-0.77}$\\
    $\Omega_m $ & $0.3075(0.3065)\pm 0.0051$ & $0.3227(0.3221)\pm 0.0095$ & $0.3162(0.3189)^{+0.0070}_{-0.010}$\\
    $w_\phi $ & \nodata & \nodata & $-0.936(-0.919)^{+0.038}_{-0.064}$ \\
    \midrule
    $\boldsymbol{\chi^2}$\textbf{ statistics} &&&\\
    $\chi^2_{\rm bf}(\Delta)$ & $2835.45$ & $2822.10(-13.5)$ & $2832.87(-2.58)$  \\
    $\chi^2_{\rm bf}/{\rm DoF}$ & $1.21$ & $1.21$ & $1.21$ \\
    Tension Level & \nodata & $3.02\sigma$ & $1.24\sigma(\text{n.s.})$\\
    \midrule
    \bottomrule
    \end{tabular}
}
\caption{
    Marginalized posteriors for flat \LCDM, $w_0w_a$CDM, and quintessence models using CMB+DESI+Union3 datasets, showing the mean (best-fit) and the $68\%$ confidence interval where the \LCDM\ parameters share the same prior across models.~We also show the best-fitting $\chi^2_{bf}(\Delta)$, where $\Delta=\chi^2_{\rm bf,model} - \chi^2_{\rm bf,\Lambda\text{CDM}}$ represents the difference between the best-fitting $\chi^2$ values with respect to \LCDM.~The level of tension with \LCDM\ are reported in the final row with ``n.s.'' indicating an insignificant tension.\label{tab
}\label{tab:marg_param}
}
\end{table*}
Our results are given in table~\ref{tab:marg_param}, with 2D contours and marginalized posteriors shown in Figure~\ref{fig:contours}.~We reproduce the DESI result that the data prefer $w_0$--$w_a$ over $\Lambda$CDM at $\sim 3\sigma$, but find no statistically significant preference for exponential quintessence.~We therefore conclude that both $\Lambda$CDM and exponential quintessence are disfavored compared with $w_0$--$w_a$.
\begin{figure}[t]
    \centering
    \includegraphics[width=\columnwidth]{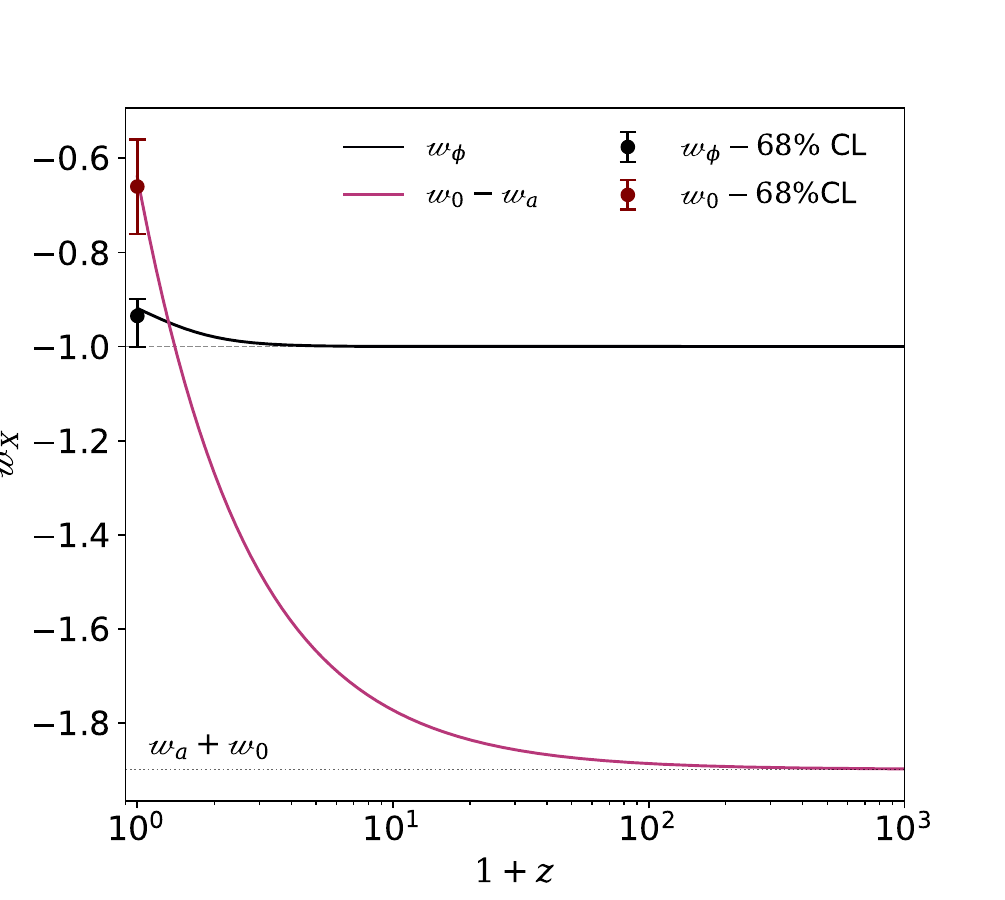}
    \caption{The equation of state for the best-fitting exponential quintessence model ($w_\phi$, black) and the CLP parameterization (red).~The dashed line corresponds to the \LCDM\ model with $w_\Lambda = -1 $, and the dotted line represents $w_0 + w_a$, which is the asymptotic EOS for the CLP parameterization for $z\gg0$.~We also show the marginalized posteriors for the EOS today for both models with the combination of CMB+DESI+Union3 datasets at the $68\%$ level.}
    \label{fig:EoS}
\end{figure}
The reason for this can be seen in figure~\ref{fig:EoS} where we plot $w(z)$ for the best-fitting $w_0$--$w_a$ and exponential quintessence models.~Both models have $w(z)>-1$ at the present time and decreasing towards more negative values in the past, but the $w_0$--$w_a$ model is able to reach $w=-1$ in a shorter time.~As discussed by DESI \cite{DESI:2024mwx} and further investigated by \cite{Colgain:2024xqj}, the DESI preference for thawing dark energy is driven by low-redshift anomalies in the supernovae and DESI BAO data.~The higher redshift DESI points are consistent with $\Lambda$CDM.~$w_0$--$w_a$ accommodates this by having $w_0>-1$ and a large negative value of $w_a$ to ensure a rapid return  $w\approx-1$.~The increasingly negative values at larger redshifts are not problematic because DE is subdominant at this time and the model behaves similarly to $\Lambda$CDM.~In contrast, the EOS for the exponential  model varies less rapidly because the field is slowly-rolling.~The EOS only tends  to $w=-1$ at higher redshifts when DE is sub-dominant, so the model is unable to accommodate each data point as well.~This suggests that quintessence potentials with sharper features e.g., hill-top or plateau models may be able to better-fit the data because they allow for more rapid variations in $w(z)$ around the onset of DE.~Indeed, reference \cite{Shlivko:2024llw} drew an identical conclusion using a different method where they determined an equivalent $w_0$-$w_a$ parameterization for three classes of quintessence models, finding that exponential models lie outside the DESI $1\sigma$ contours but that hill-top and plateau models are compatible.

Interpreting the data to identify the microphysics of dark energy remains a paramount goal of cosmology, and our results have helped to elucidate the requisite features that quintessence models must incorporate in order to accommodate the DESI data.~There are several avenues for followup investigations.~First, fitting other proposed quintessence potentials to the data would help to identify the best-fitting models;~and, second, one could look at more general scalar field models of dark energy such as coupled quintessence \cite{Amendola:1999er,Sakstein:2014aca,Sakstein:2015jca}, k-essence \cite{Chiba:1999ka,Armendariz-Picon:2000nqq,Armendariz-Picon:2000ulo}, multi-field models \cite{Cicoli:2020noz}, and modified gravity \cite{Clifton:2011jh,Joyce:2014kja,Koyama:2015vza,Ferreira:2019xrr}.~One could also go beyond scalar field models e.g., \cite{Heckman:2018mxl,Heckman:2019dsj,Heckman:2022peq}.~Our investigation suggests that any such models must allow for a sufficiently steep change in $w(z)$ around $z\sim0.5$.~In addition to the hill-top and plateau quintessence models above, we note that models such as symmetron dark energy \cite{Hinterbichler:2010es,Hinterbichler:2011ca} that use phase transitions to start a scalar field rolling possess such features, as do models where relativistic species decouple around $z\sim1$ and inject energy into a scalar such as mass-varying neutrino models \cite{Fardon:2003eh,Brookfield:2005bz,Sakstein:2019fmf,CarrilloGonzalez:2020oac}, among others.

\textbf{Note Added:}~While this manuscript was in preparation, reference \cite{Bhattacharya:2024hep}, which also studies exponential quintessence in light of the DESI data, appeared on the arXiv.~Our results agree with theirs.

\textbf{Acknowledgements:}~The technical support and advanced computing resources from University of Hawai‘i Information Technology Services – Cyberinfrastructure, funded in part by the National Science Foundation CC\* awards \#2201428 and \#2232862 are gratefully acknowledged.

\clearpage

\bibliographystyle{apsrev4-1}
\bibliography{main}

\begin{thebibliography}{50}%
\makeatletter
\providecommand \@ifxundefined [1]{%
 \@ifx{#1\undefined}
}%
\providecommand \@ifnum [1]{%
 \ifnum #1\expandafter \@firstoftwo
 \else \expandafter \@secondoftwo
 \fi
}%
\providecommand \@ifx [1]{%
 \ifx #1\expandafter \@firstoftwo
 \else \expandafter \@secondoftwo
 \fi
}%
\providecommand \natexlab [1]{#1}%
\providecommand \enquote  [1]{``#1''}%
\providecommand \bibnamefont  [1]{#1}%
\providecommand \bibfnamefont [1]{#1}%
\providecommand \citenamefont [1]{#1}%
\providecommand \href@noop [0]{\@secondoftwo}%
\providecommand \href [0]{\begingroup \@sanitize@url \@href}%
\providecommand \@href[1]{\@@startlink{#1}\@@href}%
\providecommand \@@href[1]{\endgroup#1\@@endlink}%
\providecommand \@sanitize@url [0]{\catcode `\\12\catcode `\$12\catcode `\&12\catcode `\#12\catcode `\^12\catcode `\_12\catcode `\%12\relax}%
\providecommand \@@startlink[1]{}%
\providecommand \@@endlink[0]{}%
\providecommand \url  [0]{\begingroup\@sanitize@url \@url }%
\providecommand \@url [1]{\endgroup\@href {#1}{\urlprefix }}%
\providecommand \urlprefix  [0]{URL }%
\providecommand \Eprint [0]{\href }%
\providecommand \doibase [0]{http://dx.doi.org/}%
\providecommand \selectlanguage [0]{\@gobble}%
\providecommand \bibinfo  [0]{\@secondoftwo}%
\providecommand \bibfield  [0]{\@secondoftwo}%
\providecommand \translation [1]{[#1]}%
\providecommand \BibitemOpen [0]{}%
\providecommand \bibitemStop [0]{}%
\providecommand \bibitemNoStop [0]{.\EOS\space}%
\providecommand \EOS [0]{\spacefactor3000\relax}%
\providecommand \BibitemShut  [1]{\csname bibitem#1\endcsname}%
\let\auto@bib@innerbib\@empty
\bibitem [{\citenamefont {Adame}\ \emph {et~al.}(2024)\citenamefont {Adame} \emph {et~al.}}]{DESI:2024mwx}%
  \BibitemOpen
  \bibfield  {author} {\bibinfo {author} {\bibfnamefont {A.~G.}\ \bibnamefont {Adame}} \emph {et~al.} (\bibinfo {collaboration} {DESI}),\ }\href@noop {} {\  (\bibinfo {year} {2024})},\ \Eprint {http://arxiv.org/abs/2404.03002} {arXiv:2404.03002 [astro-ph.CO]} \BibitemShut {NoStop}%
\bibitem [{\citenamefont {Brout}\ \emph {et~al.}(2022)\citenamefont {Brout} \emph {et~al.}}]{Brout:2022vxf}%
  \BibitemOpen
  \bibfield  {author} {\bibinfo {author} {\bibfnamefont {D.}~\bibnamefont {Brout}} \emph {et~al.},\ }\href {\doibase 10.3847/1538-4357/ac8e04} {\bibfield  {journal} {\bibinfo  {journal} {Astrophys. J.}\ }\textbf {\bibinfo {volume} {938}},\ \bibinfo {pages} {110} (\bibinfo {year} {2022})},\ \Eprint {http://arxiv.org/abs/2202.04077} {arXiv:2202.04077 [astro-ph.CO]} \BibitemShut {NoStop}%
\bibitem [{\citenamefont {Rubin}\ \emph {et~al.}(2023)\citenamefont {Rubin} \emph {et~al.}}]{Rubin:2023ovl}%
  \BibitemOpen
  \bibfield  {author} {\bibinfo {author} {\bibfnamefont {D.}~\bibnamefont {Rubin}} \emph {et~al.},\ }\href@noop {} {\  (\bibinfo {year} {2023})},\ \Eprint {http://arxiv.org/abs/2311.12098} {arXiv:2311.12098 [astro-ph.CO]} \BibitemShut {NoStop}%
\bibitem [{\citenamefont {Abbott}\ \emph {et~al.}(2024)\citenamefont {Abbott} \emph {et~al.}}]{DES:2024tys}%
  \BibitemOpen
  \bibfield  {author} {\bibinfo {author} {\bibfnamefont {T.~M.~C.}\ \bibnamefont {Abbott}} \emph {et~al.} (\bibinfo {collaboration} {DES}),\ }\href@noop {} {\  (\bibinfo {year} {2024})},\ \Eprint {http://arxiv.org/abs/2401.02929} {arXiv:2401.02929 [astro-ph.CO]} \BibitemShut {NoStop}%
\bibitem [{\citenamefont {Chevallier}\ and\ \citenamefont {Polarski}(2001)}]{Chevallier:2000qy}%
  \BibitemOpen
  \bibfield  {author} {\bibinfo {author} {\bibfnamefont {M.}~\bibnamefont {Chevallier}}\ and\ \bibinfo {author} {\bibfnamefont {D.}~\bibnamefont {Polarski}},\ }\href {\doibase 10.1142/S0218271801000822} {\bibfield  {journal} {\bibinfo  {journal} {Int. J. Mod. Phys. D}\ }\textbf {\bibinfo {volume} {10}},\ \bibinfo {pages} {213} (\bibinfo {year} {2001})},\ \Eprint {http://arxiv.org/abs/gr-qc/0009008} {arXiv:gr-qc/0009008} \BibitemShut {NoStop}%
\bibitem [{\citenamefont {Linder}(2003)}]{Linder:2002et}%
  \BibitemOpen
  \bibfield  {author} {\bibinfo {author} {\bibfnamefont {E.~V.}\ \bibnamefont {Linder}},\ }\href {\doibase 10.1103/PhysRevLett.90.091301} {\bibfield  {journal} {\bibinfo  {journal} {Phys. Rev. Lett.}\ }\textbf {\bibinfo {volume} {90}},\ \bibinfo {pages} {091301} (\bibinfo {year} {2003})},\ \Eprint {http://arxiv.org/abs/astro-ph/0208512} {arXiv:astro-ph/0208512} \BibitemShut {NoStop}%
\bibitem [{\citenamefont {Ratra}\ and\ \citenamefont {Peebles}(1988)}]{Ratra:1987rm}%
  \BibitemOpen
  \bibfield  {author} {\bibinfo {author} {\bibfnamefont {B.}~\bibnamefont {Ratra}}\ and\ \bibinfo {author} {\bibfnamefont {P.~J.~E.}\ \bibnamefont {Peebles}},\ }\href {\doibase 10.1103/PhysRevD.37.3406} {\bibfield  {journal} {\bibinfo  {journal} {Phys. Rev. D}\ }\textbf {\bibinfo {volume} {37}},\ \bibinfo {pages} {3406} (\bibinfo {year} {1988})}\BibitemShut {NoStop}%
\bibitem [{\citenamefont {Caldwell}\ \emph {et~al.}(1998)\citenamefont {Caldwell}, \citenamefont {Dave},\ and\ \citenamefont {Steinhardt}}]{Caldwell:1997ii}%
  \BibitemOpen
  \bibfield  {author} {\bibinfo {author} {\bibfnamefont {R.~R.}\ \bibnamefont {Caldwell}}, \bibinfo {author} {\bibfnamefont {R.}~\bibnamefont {Dave}}, \ and\ \bibinfo {author} {\bibfnamefont {P.~J.}\ \bibnamefont {Steinhardt}},\ }\href {\doibase 10.1103/PhysRevLett.80.1582} {\bibfield  {journal} {\bibinfo  {journal} {Phys. Rev. Lett.}\ }\textbf {\bibinfo {volume} {80}},\ \bibinfo {pages} {1582} (\bibinfo {year} {1998})},\ \Eprint {http://arxiv.org/abs/astro-ph/9708069} {arXiv:astro-ph/9708069} \BibitemShut {NoStop}%
\bibitem [{\citenamefont {Copeland}\ \emph {et~al.}(1998)\citenamefont {Copeland}, \citenamefont {Liddle},\ and\ \citenamefont {Wands}}]{Copeland:1997et}%
  \BibitemOpen
  \bibfield  {author} {\bibinfo {author} {\bibfnamefont {E.~J.}\ \bibnamefont {Copeland}}, \bibinfo {author} {\bibfnamefont {A.~R.}\ \bibnamefont {Liddle}}, \ and\ \bibinfo {author} {\bibfnamefont {D.}~\bibnamefont {Wands}},\ }\href {\doibase 10.1103/PhysRevD.57.4686} {\bibfield  {journal} {\bibinfo  {journal} {Phys. Rev. D}\ }\textbf {\bibinfo {volume} {57}},\ \bibinfo {pages} {4686} (\bibinfo {year} {1998})},\ \Eprint {http://arxiv.org/abs/gr-qc/9711068} {arXiv:gr-qc/9711068} \BibitemShut {NoStop}%
\bibitem [{\citenamefont {Copeland}\ \emph {et~al.}(2006)\citenamefont {Copeland}, \citenamefont {Sami},\ and\ \citenamefont {Tsujikawa}}]{Copeland:2006wr}%
  \BibitemOpen
  \bibfield  {author} {\bibinfo {author} {\bibfnamefont {E.~J.}\ \bibnamefont {Copeland}}, \bibinfo {author} {\bibfnamefont {M.}~\bibnamefont {Sami}}, \ and\ \bibinfo {author} {\bibfnamefont {S.}~\bibnamefont {Tsujikawa}},\ }\href {\doibase 10.1142/S021827180600942X} {\bibfield  {journal} {\bibinfo  {journal} {Int. J. Mod. Phys. D}\ }\textbf {\bibinfo {volume} {15}},\ \bibinfo {pages} {1753} (\bibinfo {year} {2006})},\ \Eprint {http://arxiv.org/abs/hep-th/0603057} {arXiv:hep-th/0603057} \BibitemShut {NoStop}%
\bibitem [{\citenamefont {Bahamonde}\ \emph {et~al.}(2018)\citenamefont {Bahamonde}, \citenamefont {B\"ohmer}, \citenamefont {Carloni}, \citenamefont {Copeland}, \citenamefont {Fang},\ and\ \citenamefont {Tamanini}}]{Bahamonde:2017ize}%
  \BibitemOpen
  \bibfield  {author} {\bibinfo {author} {\bibfnamefont {S.}~\bibnamefont {Bahamonde}}, \bibinfo {author} {\bibfnamefont {C.~G.}\ \bibnamefont {B\"ohmer}}, \bibinfo {author} {\bibfnamefont {S.}~\bibnamefont {Carloni}}, \bibinfo {author} {\bibfnamefont {E.~J.}\ \bibnamefont {Copeland}}, \bibinfo {author} {\bibfnamefont {W.}~\bibnamefont {Fang}}, \ and\ \bibinfo {author} {\bibfnamefont {N.}~\bibnamefont {Tamanini}},\ }\href {\doibase 10.1016/j.physrep.2018.09.001} {\bibfield  {journal} {\bibinfo  {journal} {Phys. Rept.}\ }\textbf {\bibinfo {volume} {775-777}},\ \bibinfo {pages} {1} (\bibinfo {year} {2018})},\ \Eprint {http://arxiv.org/abs/1712.03107} {arXiv:1712.03107 [gr-qc]} \BibitemShut {NoStop}%
\bibitem [{\citenamefont {Wetterich}(1995)}]{Wetterich:1994bg}%
  \BibitemOpen
  \bibfield  {author} {\bibinfo {author} {\bibfnamefont {C.}~\bibnamefont {Wetterich}},\ }\href@noop {} {\bibfield  {journal} {\bibinfo  {journal} {Astron. Astrophys.}\ }\textbf {\bibinfo {volume} {301}},\ \bibinfo {pages} {321} (\bibinfo {year} {1995})},\ \Eprint {http://arxiv.org/abs/hep-th/9408025} {arXiv:hep-th/9408025} \BibitemShut {NoStop}%
\bibitem [{\citenamefont {Binetruy}(1999)}]{Binetruy:1998rz}%
  \BibitemOpen
  \bibfield  {author} {\bibinfo {author} {\bibfnamefont {P.}~\bibnamefont {Binetruy}},\ }\href {\doibase 10.1103/PhysRevD.60.063502} {\bibfield  {journal} {\bibinfo  {journal} {Phys. Rev. D}\ }\textbf {\bibinfo {volume} {60}},\ \bibinfo {pages} {063502} (\bibinfo {year} {1999})},\ \Eprint {http://arxiv.org/abs/hep-ph/9810553} {arXiv:hep-ph/9810553} \BibitemShut {NoStop}%
\bibitem [{\citenamefont {Barreiro}\ \emph {et~al.}(2000)\citenamefont {Barreiro}, \citenamefont {Copeland},\ and\ \citenamefont {Nunes}}]{Barreiro:1999zs}%
  \BibitemOpen
  \bibfield  {author} {\bibinfo {author} {\bibfnamefont {T.}~\bibnamefont {Barreiro}}, \bibinfo {author} {\bibfnamefont {E.~J.}\ \bibnamefont {Copeland}}, \ and\ \bibinfo {author} {\bibfnamefont {N.~J.}\ \bibnamefont {Nunes}},\ }\href {\doibase 10.1103/PhysRevD.61.127301} {\bibfield  {journal} {\bibinfo  {journal} {Phys. Rev. D}\ }\textbf {\bibinfo {volume} {61}},\ \bibinfo {pages} {127301} (\bibinfo {year} {2000})},\ \Eprint {http://arxiv.org/abs/astro-ph/9910214} {arXiv:astro-ph/9910214} \BibitemShut {NoStop}%
\bibitem [{\citenamefont {Ramadan}\ \emph {et~al.}(2024)\citenamefont {Ramadan}, \citenamefont {Karwal},\ and\ \citenamefont {Sakstein}}]{Ramadan:2023ivw}%
  \BibitemOpen
  \bibfield  {author} {\bibinfo {author} {\bibfnamefont {O.~F.}\ \bibnamefont {Ramadan}}, \bibinfo {author} {\bibfnamefont {T.}~\bibnamefont {Karwal}}, \ and\ \bibinfo {author} {\bibfnamefont {J.}~\bibnamefont {Sakstein}},\ }\href {\doibase 10.1103/PhysRevD.109.063525} {\bibfield  {journal} {\bibinfo  {journal} {Phys. Rev. D}\ }\textbf {\bibinfo {volume} {109}},\ \bibinfo {pages} {063525} (\bibinfo {year} {2024})},\ \Eprint {http://arxiv.org/abs/2309.08082} {arXiv:2309.08082 [astro-ph.CO]} \BibitemShut {NoStop}%
\bibitem [{\citenamefont {Aghanim}\ \emph {et~al.}(2020)\citenamefont {Aghanim} \emph {et~al.}}]{Planck:2018vyg}%
  \BibitemOpen
  \bibfield  {author} {\bibinfo {author} {\bibfnamefont {N.}~\bibnamefont {Aghanim}} \emph {et~al.} (\bibinfo {collaboration} {Planck}),\ }\href {\doibase 10.1051/0004-6361/201833910} {\bibfield  {journal} {\bibinfo  {journal} {Astron. Astrophys.}\ }\textbf {\bibinfo {volume} {641}},\ \bibinfo {pages} {A6} (\bibinfo {year} {2020})},\ \bibinfo {note} {[Erratum: Astron.Astrophys. 652, C4 (2021)]},\ \Eprint {http://arxiv.org/abs/1807.06209} {arXiv:1807.06209 [astro-ph.CO]} \BibitemShut {NoStop}%
\bibitem [{\citenamefont {Akrami}\ \emph {et~al.}(2020)\citenamefont {Akrami} \emph {et~al.}}]{Planck:2020olo}%
  \BibitemOpen
  \bibfield  {author} {\bibinfo {author} {\bibfnamefont {Y.}~\bibnamefont {Akrami}} \emph {et~al.} (\bibinfo {collaboration} {Planck}),\ }\href {\doibase 10.1051/0004-6361/202038073} {\bibfield  {journal} {\bibinfo  {journal} {Astron. Astrophys.}\ }\textbf {\bibinfo {volume} {643}},\ \bibinfo {pages} {A42} (\bibinfo {year} {2020})},\ \Eprint {http://arxiv.org/abs/2007.04997} {arXiv:2007.04997 [astro-ph.CO]} \BibitemShut {NoStop}%
\bibitem [{\citenamefont {Carron}\ \emph {et~al.}(2022)\citenamefont {Carron}, \citenamefont {Mirmelstein},\ and\ \citenamefont {Lewis}}]{Carron:2022eyg}%
  \BibitemOpen
  \bibfield  {author} {\bibinfo {author} {\bibfnamefont {J.}~\bibnamefont {Carron}}, \bibinfo {author} {\bibfnamefont {M.}~\bibnamefont {Mirmelstein}}, \ and\ \bibinfo {author} {\bibfnamefont {A.}~\bibnamefont {Lewis}},\ }\href {\doibase 10.1088/1475-7516/2022/09/039} {\bibfield  {journal} {\bibinfo  {journal} {JCAP}\ }\textbf {\bibinfo {volume} {09}},\ \bibinfo {pages} {039} (\bibinfo {year} {2022})},\ \Eprint {http://arxiv.org/abs/2206.07773} {arXiv:2206.07773 [astro-ph.CO]} \BibitemShut {NoStop}%
\bibitem [{\citenamefont {Qu}\ \emph {et~al.}(2024)\citenamefont {Qu} \emph {et~al.}}]{ACT:2023dou}%
  \BibitemOpen
  \bibfield  {author} {\bibinfo {author} {\bibfnamefont {F.~J.}\ \bibnamefont {Qu}} \emph {et~al.} (\bibinfo {collaboration} {ACT}),\ }\href {\doibase 10.3847/1538-4357/acfe06} {\bibfield  {journal} {\bibinfo  {journal} {Astrophys. J.}\ }\textbf {\bibinfo {volume} {962}},\ \bibinfo {pages} {112} (\bibinfo {year} {2024})},\ \Eprint {http://arxiv.org/abs/2304.05202} {arXiv:2304.05202 [astro-ph.CO]} \BibitemShut {NoStop}%
\bibitem [{\citenamefont {Madhavacheril}\ \emph {et~al.}(2024)\citenamefont {Madhavacheril} \emph {et~al.}}]{ACT:2023kun}%
  \BibitemOpen
  \bibfield  {author} {\bibinfo {author} {\bibfnamefont {M.~S.}\ \bibnamefont {Madhavacheril}} \emph {et~al.} (\bibinfo {collaboration} {ACT}),\ }\href {\doibase 10.3847/1538-4357/acff5f} {\bibfield  {journal} {\bibinfo  {journal} {Astrophys. J.}\ }\textbf {\bibinfo {volume} {962}},\ \bibinfo {pages} {113} (\bibinfo {year} {2024})},\ \Eprint {http://arxiv.org/abs/2304.05203} {arXiv:2304.05203 [astro-ph.CO]} \BibitemShut {NoStop}%
\bibitem [{\citenamefont {Blas}\ \emph {et~al.}(2011)\citenamefont {Blas}, \citenamefont {Lesgourgues},\ and\ \citenamefont {Tram}}]{Blas:2011rf}%
  \BibitemOpen
  \bibfield  {author} {\bibinfo {author} {\bibfnamefont {D.}~\bibnamefont {Blas}}, \bibinfo {author} {\bibfnamefont {J.}~\bibnamefont {Lesgourgues}}, \ and\ \bibinfo {author} {\bibfnamefont {T.}~\bibnamefont {Tram}},\ }\href {\doibase 10.1088/1475-7516/2011/07/034} {\bibfield  {journal} {\bibinfo  {journal} {JCAP}\ }\textbf {\bibinfo {volume} {07}},\ \bibinfo {pages} {034} (\bibinfo {year} {2011})},\ \Eprint {http://arxiv.org/abs/1104.2933} {arXiv:1104.2933 [astro-ph.CO]} \BibitemShut {NoStop}%
\bibitem [{\citenamefont {Lesgourgues}(2011)}]{Lesgourgues:2011re}%
  \BibitemOpen
  \bibfield  {author} {\bibinfo {author} {\bibfnamefont {J.}~\bibnamefont {Lesgourgues}},\ }\href@noop {} {\  (\bibinfo {year} {2011})},\ \bibinfo {note} {arXiv:1104.2932}\BibitemShut {NoStop}%
\bibitem [{\citenamefont {Torrado}\ and\ \citenamefont {Lewis}(2021)}]{Torrado:2020dgo}%
  \BibitemOpen
  \bibfield  {author} {\bibinfo {author} {\bibfnamefont {J.}~\bibnamefont {Torrado}}\ and\ \bibinfo {author} {\bibfnamefont {A.}~\bibnamefont {Lewis}},\ }\href {\doibase 10.1088/1475-7516/2021/05/057} {\bibfield  {journal} {\bibinfo  {journal} {JCAP}\ }\textbf {\bibinfo {volume} {05}},\ \bibinfo {pages} {057} (\bibinfo {year} {2021})},\ \bibinfo {note} {arXiv:2005.05290}\BibitemShut {NoStop}%
\bibitem [{\citenamefont {Hastings}(1970)}]{hastings1970monte}%
  \BibitemOpen
  \bibfield  {author} {\bibinfo {author} {\bibfnamefont {W.~K.}\ \bibnamefont {Hastings}},\ }\href@noop {} {\bibfield  {journal} {\bibinfo  {journal} {Biometrika}\ }\textbf {\bibinfo {volume} {57}},\ \bibinfo {pages} {97} (\bibinfo {year} {1970})}\BibitemShut {NoStop}%
\bibitem [{\citenamefont {Metropolis}\ \emph {et~al.}(1953)\citenamefont {Metropolis}, \citenamefont {Rosenbluth}, \citenamefont {Rosenbluth}, \citenamefont {Teller},\ and\ \citenamefont {Teller}}]{metropolis1953equation}%
  \BibitemOpen
  \bibfield  {author} {\bibinfo {author} {\bibfnamefont {N.}~\bibnamefont {Metropolis}}, \bibinfo {author} {\bibfnamefont {A.~W.}\ \bibnamefont {Rosenbluth}}, \bibinfo {author} {\bibfnamefont {M.~N.}\ \bibnamefont {Rosenbluth}}, \bibinfo {author} {\bibfnamefont {A.~H.}\ \bibnamefont {Teller}}, \ and\ \bibinfo {author} {\bibfnamefont {E.}~\bibnamefont {Teller}},\ }\href@noop {} {\bibfield  {journal} {\bibinfo  {journal} {The journal of chemical physics}\ }\textbf {\bibinfo {volume} {21}},\ \bibinfo {pages} {1087} (\bibinfo {year} {1953})}\BibitemShut {NoStop}%
\bibitem [{\citenamefont {Gelman}\ and\ \citenamefont {Rubin}(1992)}]{GelmanRubin1992}%
  \BibitemOpen
  \bibfield  {author} {\bibinfo {author} {\bibfnamefont {A.}~\bibnamefont {Gelman}}\ and\ \bibinfo {author} {\bibfnamefont {D.~B.}\ \bibnamefont {Rubin}},\ }\href {\doibase 10.1214/ss/1177011136} {\bibfield  {journal} {\bibinfo  {journal} {Statist. Sci.}\ }\textbf {\bibinfo {volume} {7}},\ \bibinfo {pages} {457} (\bibinfo {year} {1992})}\BibitemShut {NoStop}%
\bibitem [{\citenamefont {Lewis}(2019)}]{Lewis:2019xzd}%
  \BibitemOpen
  \bibfield  {author} {\bibinfo {author} {\bibfnamefont {A.}~\bibnamefont {Lewis}},\ }\href@noop {} {\  (\bibinfo {year} {2019})},\ \Eprint {http://arxiv.org/abs/1910.13970} {arXiv:1910.13970 [astro-ph.IM]} \BibitemShut {NoStop}%
\bibitem [{\citenamefont {Colg\'ain}\ \emph {et~al.}(2024)\citenamefont {Colg\'ain}, \citenamefont {Dainotti}, \citenamefont {Capozziello}, \citenamefont {Pourojaghi}, \citenamefont {Sheikh-Jabbari},\ and\ \citenamefont {Stojkovic}}]{Colgain:2024xqj}%
  \BibitemOpen
  \bibfield  {author} {\bibinfo {author} {\bibfnamefont {E.~O.}\ \bibnamefont {Colg\'ain}}, \bibinfo {author} {\bibfnamefont {M.~G.}\ \bibnamefont {Dainotti}}, \bibinfo {author} {\bibfnamefont {S.}~\bibnamefont {Capozziello}}, \bibinfo {author} {\bibfnamefont {S.}~\bibnamefont {Pourojaghi}}, \bibinfo {author} {\bibfnamefont {M.~M.}\ \bibnamefont {Sheikh-Jabbari}}, \ and\ \bibinfo {author} {\bibfnamefont {D.}~\bibnamefont {Stojkovic}},\ }\href@noop {} {\  (\bibinfo {year} {2024})},\ \Eprint {http://arxiv.org/abs/2404.08633} {arXiv:2404.08633 [astro-ph.CO]} \BibitemShut {NoStop}%
\bibitem [{\citenamefont {Shlivko}\ and\ \citenamefont {Steinhardt}(2024)}]{Shlivko:2024llw}%
  \BibitemOpen
  \bibfield  {author} {\bibinfo {author} {\bibfnamefont {D.}~\bibnamefont {Shlivko}}\ and\ \bibinfo {author} {\bibfnamefont {P.}~\bibnamefont {Steinhardt}},\ }\href@noop {} {\  (\bibinfo {year} {2024})},\ \Eprint {http://arxiv.org/abs/2405.03933} {arXiv:2405.03933 [astro-ph.CO]} \BibitemShut {NoStop}%
\bibitem [{\citenamefont {Amendola}(2000)}]{Amendola:1999er}%
  \BibitemOpen
  \bibfield  {author} {\bibinfo {author} {\bibfnamefont {L.}~\bibnamefont {Amendola}},\ }\href {\doibase 10.1103/PhysRevD.62.043511} {\bibfield  {journal} {\bibinfo  {journal} {Phys. Rev. D}\ }\textbf {\bibinfo {volume} {62}},\ \bibinfo {pages} {043511} (\bibinfo {year} {2000})},\ \Eprint {http://arxiv.org/abs/astro-ph/9908023} {arXiv:astro-ph/9908023} \BibitemShut {NoStop}%
\bibitem [{\citenamefont {Sakstein}(2015)}]{Sakstein:2014aca}%
  \BibitemOpen
  \bibfield  {author} {\bibinfo {author} {\bibfnamefont {J.}~\bibnamefont {Sakstein}},\ }\href {\doibase 10.1103/PhysRevD.91.024036} {\bibfield  {journal} {\bibinfo  {journal} {Phys. Rev. D}\ }\textbf {\bibinfo {volume} {91}},\ \bibinfo {pages} {024036} (\bibinfo {year} {2015})},\ \Eprint {http://arxiv.org/abs/1409.7296} {arXiv:1409.7296 [astro-ph.CO]} \BibitemShut {NoStop}%
\bibitem [{\citenamefont {Sakstein}\ and\ \citenamefont {Verner}(2015)}]{Sakstein:2015jca}%
  \BibitemOpen
  \bibfield  {author} {\bibinfo {author} {\bibfnamefont {J.}~\bibnamefont {Sakstein}}\ and\ \bibinfo {author} {\bibfnamefont {S.}~\bibnamefont {Verner}},\ }\href {\doibase 10.1103/PhysRevD.92.123005} {\bibfield  {journal} {\bibinfo  {journal} {Phys. Rev. D}\ }\textbf {\bibinfo {volume} {92}},\ \bibinfo {pages} {123005} (\bibinfo {year} {2015})},\ \Eprint {http://arxiv.org/abs/1509.05679} {arXiv:1509.05679 [gr-qc]} \BibitemShut {NoStop}%
\bibitem [{\citenamefont {Chiba}\ \emph {et~al.}(2000)\citenamefont {Chiba}, \citenamefont {Okabe},\ and\ \citenamefont {Yamaguchi}}]{Chiba:1999ka}%
  \BibitemOpen
  \bibfield  {author} {\bibinfo {author} {\bibfnamefont {T.}~\bibnamefont {Chiba}}, \bibinfo {author} {\bibfnamefont {T.}~\bibnamefont {Okabe}}, \ and\ \bibinfo {author} {\bibfnamefont {M.}~\bibnamefont {Yamaguchi}},\ }\href {\doibase 10.1103/PhysRevD.62.023511} {\bibfield  {journal} {\bibinfo  {journal} {Phys. Rev. D}\ }\textbf {\bibinfo {volume} {62}},\ \bibinfo {pages} {023511} (\bibinfo {year} {2000})},\ \Eprint {http://arxiv.org/abs/astro-ph/9912463} {arXiv:astro-ph/9912463} \BibitemShut {NoStop}%
\bibitem [{\citenamefont {Armendariz-Picon}\ \emph {et~al.}(2000)\citenamefont {Armendariz-Picon}, \citenamefont {Mukhanov},\ and\ \citenamefont {Steinhardt}}]{Armendariz-Picon:2000nqq}%
  \BibitemOpen
  \bibfield  {author} {\bibinfo {author} {\bibfnamefont {C.}~\bibnamefont {Armendariz-Picon}}, \bibinfo {author} {\bibfnamefont {V.~F.}\ \bibnamefont {Mukhanov}}, \ and\ \bibinfo {author} {\bibfnamefont {P.~J.}\ \bibnamefont {Steinhardt}},\ }\href {\doibase 10.1103/PhysRevLett.85.4438} {\bibfield  {journal} {\bibinfo  {journal} {Phys. Rev. Lett.}\ }\textbf {\bibinfo {volume} {85}},\ \bibinfo {pages} {4438} (\bibinfo {year} {2000})},\ \Eprint {http://arxiv.org/abs/astro-ph/0004134} {arXiv:astro-ph/0004134} \BibitemShut {NoStop}%
\bibitem [{\citenamefont {Armendariz-Picon}\ \emph {et~al.}(2001)\citenamefont {Armendariz-Picon}, \citenamefont {Mukhanov},\ and\ \citenamefont {Steinhardt}}]{Armendariz-Picon:2000ulo}%
  \BibitemOpen
  \bibfield  {author} {\bibinfo {author} {\bibfnamefont {C.}~\bibnamefont {Armendariz-Picon}}, \bibinfo {author} {\bibfnamefont {V.~F.}\ \bibnamefont {Mukhanov}}, \ and\ \bibinfo {author} {\bibfnamefont {P.~J.}\ \bibnamefont {Steinhardt}},\ }\href {\doibase 10.1103/PhysRevD.63.103510} {\bibfield  {journal} {\bibinfo  {journal} {Phys. Rev. D}\ }\textbf {\bibinfo {volume} {63}},\ \bibinfo {pages} {103510} (\bibinfo {year} {2001})},\ \Eprint {http://arxiv.org/abs/astro-ph/0006373} {arXiv:astro-ph/0006373} \BibitemShut {NoStop}%
\bibitem [{\citenamefont {Cicoli}\ \emph {et~al.}(2020)\citenamefont {Cicoli}, \citenamefont {Dibitetto},\ and\ \citenamefont {Pedro}}]{Cicoli:2020noz}%
  \BibitemOpen
  \bibfield  {author} {\bibinfo {author} {\bibfnamefont {M.}~\bibnamefont {Cicoli}}, \bibinfo {author} {\bibfnamefont {G.}~\bibnamefont {Dibitetto}}, \ and\ \bibinfo {author} {\bibfnamefont {F.~G.}\ \bibnamefont {Pedro}},\ }\href {\doibase 10.1007/JHEP10(2020)035} {\bibfield  {journal} {\bibinfo  {journal} {JHEP}\ }\textbf {\bibinfo {volume} {10}},\ \bibinfo {pages} {035} (\bibinfo {year} {2020})},\ \Eprint {http://arxiv.org/abs/2007.11011} {arXiv:2007.11011 [hep-th]} \BibitemShut {NoStop}%
\bibitem [{\citenamefont {Clifton}\ \emph {et~al.}(2012)\citenamefont {Clifton}, \citenamefont {Ferreira}, \citenamefont {Padilla},\ and\ \citenamefont {Skordis}}]{Clifton:2011jh}%
  \BibitemOpen
  \bibfield  {author} {\bibinfo {author} {\bibfnamefont {T.}~\bibnamefont {Clifton}}, \bibinfo {author} {\bibfnamefont {P.~G.}\ \bibnamefont {Ferreira}}, \bibinfo {author} {\bibfnamefont {A.}~\bibnamefont {Padilla}}, \ and\ \bibinfo {author} {\bibfnamefont {C.}~\bibnamefont {Skordis}},\ }\href {\doibase 10.1016/j.physrep.2012.01.001} {\bibfield  {journal} {\bibinfo  {journal} {Phys. Rept.}\ }\textbf {\bibinfo {volume} {513}},\ \bibinfo {pages} {1} (\bibinfo {year} {2012})},\ \Eprint {http://arxiv.org/abs/1106.2476} {arXiv:1106.2476 [astro-ph.CO]} \BibitemShut {NoStop}%
\bibitem [{\citenamefont {Joyce}\ \emph {et~al.}(2015)\citenamefont {Joyce}, \citenamefont {Jain}, \citenamefont {Khoury},\ and\ \citenamefont {Trodden}}]{Joyce:2014kja}%
  \BibitemOpen
  \bibfield  {author} {\bibinfo {author} {\bibfnamefont {A.}~\bibnamefont {Joyce}}, \bibinfo {author} {\bibfnamefont {B.}~\bibnamefont {Jain}}, \bibinfo {author} {\bibfnamefont {J.}~\bibnamefont {Khoury}}, \ and\ \bibinfo {author} {\bibfnamefont {M.}~\bibnamefont {Trodden}},\ }\href {\doibase 10.1016/j.physrep.2014.12.002} {\bibfield  {journal} {\bibinfo  {journal} {Phys. Rept.}\ }\textbf {\bibinfo {volume} {568}},\ \bibinfo {pages} {1} (\bibinfo {year} {2015})},\ \Eprint {http://arxiv.org/abs/1407.0059} {arXiv:1407.0059 [astro-ph.CO]} \BibitemShut {NoStop}%
\bibitem [{\citenamefont {Koyama}(2016)}]{Koyama:2015vza}%
  \BibitemOpen
  \bibfield  {author} {\bibinfo {author} {\bibfnamefont {K.}~\bibnamefont {Koyama}},\ }\href {\doibase 10.1088/0034-4885/79/4/046902} {\bibfield  {journal} {\bibinfo  {journal} {Rept. Prog. Phys.}\ }\textbf {\bibinfo {volume} {79}},\ \bibinfo {pages} {046902} (\bibinfo {year} {2016})},\ \Eprint {http://arxiv.org/abs/1504.04623} {arXiv:1504.04623 [astro-ph.CO]} \BibitemShut {NoStop}%
\bibitem [{\citenamefont {Ferreira}(2019)}]{Ferreira:2019xrr}%
  \BibitemOpen
  \bibfield  {author} {\bibinfo {author} {\bibfnamefont {P.~G.}\ \bibnamefont {Ferreira}},\ }\href {\doibase 10.1146/annurev-astro-091918-104423} {\bibfield  {journal} {\bibinfo  {journal} {Ann. Rev. Astron. Astrophys.}\ }\textbf {\bibinfo {volume} {57}},\ \bibinfo {pages} {335} (\bibinfo {year} {2019})},\ \Eprint {http://arxiv.org/abs/1902.10503} {arXiv:1902.10503 [astro-ph.CO]} \BibitemShut {NoStop}%
\bibitem [{\citenamefont {Heckman}\ \emph {et~al.}(2019{\natexlab{a}})\citenamefont {Heckman}, \citenamefont {Lawrie}, \citenamefont {Lin},\ and\ \citenamefont {Zoccarato}}]{Heckman:2018mxl}%
  \BibitemOpen
  \bibfield  {author} {\bibinfo {author} {\bibfnamefont {J.~J.}\ \bibnamefont {Heckman}}, \bibinfo {author} {\bibfnamefont {C.}~\bibnamefont {Lawrie}}, \bibinfo {author} {\bibfnamefont {L.}~\bibnamefont {Lin}}, \ and\ \bibinfo {author} {\bibfnamefont {G.}~\bibnamefont {Zoccarato}},\ }\href {\doibase 10.1002/prop.201900057} {\bibfield  {journal} {\bibinfo  {journal} {Fortsch. Phys.}\ }\textbf {\bibinfo {volume} {67}},\ \bibinfo {pages} {1900057} (\bibinfo {year} {2019}{\natexlab{a}})},\ \Eprint {http://arxiv.org/abs/1811.01959} {arXiv:1811.01959 [hep-th]} \BibitemShut {NoStop}%
\bibitem [{\citenamefont {Heckman}\ \emph {et~al.}(2019{\natexlab{b}})\citenamefont {Heckman}, \citenamefont {Lawrie}, \citenamefont {Lin}, \citenamefont {Sakstein},\ and\ \citenamefont {Zoccarato}}]{Heckman:2019dsj}%
  \BibitemOpen
  \bibfield  {author} {\bibinfo {author} {\bibfnamefont {J.~J.}\ \bibnamefont {Heckman}}, \bibinfo {author} {\bibfnamefont {C.}~\bibnamefont {Lawrie}}, \bibinfo {author} {\bibfnamefont {L.}~\bibnamefont {Lin}}, \bibinfo {author} {\bibfnamefont {J.}~\bibnamefont {Sakstein}}, \ and\ \bibinfo {author} {\bibfnamefont {G.}~\bibnamefont {Zoccarato}},\ }\href {\doibase 10.1002/prop.201900071} {\bibfield  {journal} {\bibinfo  {journal} {Fortsch. Phys.}\ }\textbf {\bibinfo {volume} {67}},\ \bibinfo {pages} {1900071} (\bibinfo {year} {2019}{\natexlab{b}})},\ \Eprint {http://arxiv.org/abs/1901.10489} {arXiv:1901.10489 [hep-th]} \BibitemShut {NoStop}%
\bibitem [{\citenamefont {Heckman}\ \emph {et~al.}(2022)\citenamefont {Heckman}, \citenamefont {Joyce}, \citenamefont {Sakstein},\ and\ \citenamefont {Trodden}}]{Heckman:2022peq}%
  \BibitemOpen
  \bibfield  {author} {\bibinfo {author} {\bibfnamefont {J.~J.}\ \bibnamefont {Heckman}}, \bibinfo {author} {\bibfnamefont {A.}~\bibnamefont {Joyce}}, \bibinfo {author} {\bibfnamefont {J.}~\bibnamefont {Sakstein}}, \ and\ \bibinfo {author} {\bibfnamefont {M.}~\bibnamefont {Trodden}},\ }\href {\doibase 10.1142/S0217751X22502013} {\bibfield  {journal} {\bibinfo  {journal} {Int. J. Mod. Phys. A}\ }\textbf {\bibinfo {volume} {37}},\ \bibinfo {pages} {2250201} (\bibinfo {year} {2022})},\ \Eprint {http://arxiv.org/abs/2208.02267} {arXiv:2208.02267 [hep-th]} \BibitemShut {NoStop}%
\bibitem [{\citenamefont {Hinterbichler}\ and\ \citenamefont {Khoury}(2010)}]{Hinterbichler:2010es}%
  \BibitemOpen
  \bibfield  {author} {\bibinfo {author} {\bibfnamefont {K.}~\bibnamefont {Hinterbichler}}\ and\ \bibinfo {author} {\bibfnamefont {J.}~\bibnamefont {Khoury}},\ }\href {\doibase 10.1103/PhysRevLett.104.231301} {\bibfield  {journal} {\bibinfo  {journal} {Phys. Rev. Lett.}\ }\textbf {\bibinfo {volume} {104}},\ \bibinfo {pages} {231301} (\bibinfo {year} {2010})},\ \Eprint {http://arxiv.org/abs/1001.4525} {arXiv:1001.4525 [hep-th]} \BibitemShut {NoStop}%
\bibitem [{\citenamefont {Hinterbichler}\ \emph {et~al.}(2011)\citenamefont {Hinterbichler}, \citenamefont {Khoury}, \citenamefont {Levy},\ and\ \citenamefont {Matas}}]{Hinterbichler:2011ca}%
  \BibitemOpen
  \bibfield  {author} {\bibinfo {author} {\bibfnamefont {K.}~\bibnamefont {Hinterbichler}}, \bibinfo {author} {\bibfnamefont {J.}~\bibnamefont {Khoury}}, \bibinfo {author} {\bibfnamefont {A.}~\bibnamefont {Levy}}, \ and\ \bibinfo {author} {\bibfnamefont {A.}~\bibnamefont {Matas}},\ }\href {\doibase 10.1103/PhysRevD.84.103521} {\bibfield  {journal} {\bibinfo  {journal} {Phys. Rev. D}\ }\textbf {\bibinfo {volume} {84}},\ \bibinfo {pages} {103521} (\bibinfo {year} {2011})},\ \Eprint {http://arxiv.org/abs/1107.2112} {arXiv:1107.2112 [astro-ph.CO]} \BibitemShut {NoStop}%
\bibitem [{\citenamefont {Fardon}\ \emph {et~al.}(2004)\citenamefont {Fardon}, \citenamefont {Nelson},\ and\ \citenamefont {Weiner}}]{Fardon:2003eh}%
  \BibitemOpen
  \bibfield  {author} {\bibinfo {author} {\bibfnamefont {R.}~\bibnamefont {Fardon}}, \bibinfo {author} {\bibfnamefont {A.~E.}\ \bibnamefont {Nelson}}, \ and\ \bibinfo {author} {\bibfnamefont {N.}~\bibnamefont {Weiner}},\ }\href {\doibase 10.1088/1475-7516/2004/10/005} {\bibfield  {journal} {\bibinfo  {journal} {JCAP}\ }\textbf {\bibinfo {volume} {10}},\ \bibinfo {pages} {005} (\bibinfo {year} {2004})},\ \Eprint {http://arxiv.org/abs/astro-ph/0309800} {arXiv:astro-ph/0309800} \BibitemShut {NoStop}%
\bibitem [{\citenamefont {Brookfield}\ \emph {et~al.}(2006)\citenamefont {Brookfield}, \citenamefont {van~de Bruck}, \citenamefont {Mota},\ and\ \citenamefont {Tocchini-Valentini}}]{Brookfield:2005bz}%
  \BibitemOpen
  \bibfield  {author} {\bibinfo {author} {\bibfnamefont {A.~W.}\ \bibnamefont {Brookfield}}, \bibinfo {author} {\bibfnamefont {C.}~\bibnamefont {van~de Bruck}}, \bibinfo {author} {\bibfnamefont {D.~F.}\ \bibnamefont {Mota}}, \ and\ \bibinfo {author} {\bibfnamefont {D.}~\bibnamefont {Tocchini-Valentini}},\ }\href {\doibase 10.1103/PhysRevD.73.083515} {\bibfield  {journal} {\bibinfo  {journal} {Phys. Rev. D}\ }\textbf {\bibinfo {volume} {73}},\ \bibinfo {pages} {083515} (\bibinfo {year} {2006})},\ \bibinfo {note} {[Erratum: Phys.Rev.D 76, 049901 (2007)]},\ \Eprint {http://arxiv.org/abs/astro-ph/0512367} {arXiv:astro-ph/0512367} \BibitemShut {NoStop}%
\bibitem [{\citenamefont {Sakstein}\ and\ \citenamefont {Trodden}(2020)}]{Sakstein:2019fmf}%
  \BibitemOpen
  \bibfield  {author} {\bibinfo {author} {\bibfnamefont {J.}~\bibnamefont {Sakstein}}\ and\ \bibinfo {author} {\bibfnamefont {M.}~\bibnamefont {Trodden}},\ }\href {\doibase 10.1103/PhysRevLett.124.161301} {\bibfield  {journal} {\bibinfo  {journal} {Phys. Rev. Lett.}\ }\textbf {\bibinfo {volume} {124}},\ \bibinfo {pages} {161301} (\bibinfo {year} {2020})},\ \Eprint {http://arxiv.org/abs/1911.11760} {arXiv:1911.11760 [astro-ph.CO]} \BibitemShut {NoStop}%
\bibitem [{\citenamefont {Carrillo~Gonz\'alez}\ \emph {et~al.}(2021)\citenamefont {Carrillo~Gonz\'alez}, \citenamefont {Liang}, \citenamefont {Sakstein},\ and\ \citenamefont {Trodden}}]{CarrilloGonzalez:2020oac}%
  \BibitemOpen
  \bibfield  {author} {\bibinfo {author} {\bibfnamefont {M.}~\bibnamefont {Carrillo~Gonz\'alez}}, \bibinfo {author} {\bibfnamefont {Q.}~\bibnamefont {Liang}}, \bibinfo {author} {\bibfnamefont {J.}~\bibnamefont {Sakstein}}, \ and\ \bibinfo {author} {\bibfnamefont {M.}~\bibnamefont {Trodden}},\ }\href {\doibase 10.1088/1475-7516/2021/04/063} {\bibfield  {journal} {\bibinfo  {journal} {JCAP}\ }\textbf {\bibinfo {volume} {04}},\ \bibinfo {pages} {063} (\bibinfo {year} {2021})},\ \Eprint {http://arxiv.org/abs/2011.09895} {arXiv:2011.09895 [astro-ph.CO]} \BibitemShut {NoStop}%
\bibitem [{\citenamefont {Bhattacharya}\ \emph {et~al.}(2024)\citenamefont {Bhattacharya}, \citenamefont {Borghetto}, \citenamefont {Malhotra}, \citenamefont {Parameswaran}, \citenamefont {Tasinato},\ and\ \citenamefont {Zavala}}]{Bhattacharya:2024hep}%
  \BibitemOpen
  \bibfield  {author} {\bibinfo {author} {\bibfnamefont {S.}~\bibnamefont {Bhattacharya}}, \bibinfo {author} {\bibfnamefont {G.}~\bibnamefont {Borghetto}}, \bibinfo {author} {\bibfnamefont {A.}~\bibnamefont {Malhotra}}, \bibinfo {author} {\bibfnamefont {S.}~\bibnamefont {Parameswaran}}, \bibinfo {author} {\bibfnamefont {G.}~\bibnamefont {Tasinato}}, \ and\ \bibinfo {author} {\bibfnamefont {I.}~\bibnamefont {Zavala}},\ }\href@noop {} {\  (\bibinfo {year} {2024})},\ \Eprint {http://arxiv.org/abs/2405.17396} {arXiv:2405.17396 [astro-ph.CO]} \BibitemShut {NoStop}%
\end{thebibliography}%

\end{document}